\documentstyle[12pt]{article}
\def\beq{\begin{equation}}
\def\eeq{\end{equation}}
\def\bea{\begin{eqnarray}}
\def\eea{\end{eqnarray}}
\begin {document}
\begin{titlepage}
October 1994 \hfill DESY 94-? \\
\begin{flushright}
HU Berlin-IEP-94/21 \\
\end{flushright}
\mbox{ }  \hfill hepth@xxx/9410084
\vspace{6ex}
\Large
\begin {center}
\bf{On gravitational dressing of renormalization group $\beta$-functions
beyond lowest order of perturbation theory}
\end {center}
\large
\vspace{3ex}
\begin{center}
H. Dorn
\footnote{e-mail: dorn@ifh.de}
\end{center}
\normalsize
\it
\vspace{3ex}
\begin{center}
Humboldt--Universit\"at zu Berlin \\
Institut f\"ur Physik, Theorie der Elementarteilchen \\
Invalidenstra\ss e 110, D-10115 Berlin
\end{center}
\vspace{6 ex }
\rm
\begin{center}
\bf{Abstract}
\end{center}
\vspace{3ex}
Based on considerations in conformal gauge I derive up to $nextleading$
order a relation between
the coefficients of $\beta $-functions in 2D renormalizable field
theories before and after coupling to quantized gravity. The result
implies a coupling constant dependence of the ratio of both
$\beta $-functions beyond leading order.
\end {titlepage}
\newpage
\setcounter{page}{1}
\pagestyle{plain}
\section {Introduction}
The coupling of 2D conformal field theories to quantized 2D gravity
(gravitational
dressing) is well understood \cite{KPZ,David,DK,DO1}, at least for central
charges
$c\leq 1$ or $c\geq 25$. Only recently there has been some progress in the
general
discussion of gravitational dressing of the larger class of renormalizable 2D
field
theories. One considers a 2D theory described by the action
\beq
S~=~S_{c}~+~\sum_{i} g_{i}\int V_{i}~ d^{2}z~,
\label{1}
\eeq
where $S_{c}$ is the action of a conformal field theory with central charge
$c$,
$V_{i}$ a set of marginal operators with respect to $S_{c}$ which is closed
under renormalization and $g_{i}$ dimensionless couplings. Then it has been
shown
\cite{Schmidh,KKP,AG,TKS} that the gravitational dressed RG $\beta$-functions
$\bar{\beta _{i}}(g)$ in $lowest~order$ are related to the original
$\beta$-functions $\beta _{i}(g)$ corresponding to the action $S$ by the
universal formula
\beq
\bar{\beta _{i}}(g)~=~\frac{2}{\alpha Q}~\beta_{i}(g)~.
\label{2}
\eeq
$\alpha$ and $Q$ are fixed by the central charge $c$ of the unperturbed
theory $S_{c}$
\beq
Q~=~\sqrt {\frac{25-c}{3}}
\label{3}
\eeq
\beq
\alpha ~=~\frac{Q-\sqrt{Q^{2}-8}}{2}~.
\label{4}
\eeq
In its generalization to the case of an infinite number of couplings i.e.
to generalized $\sigma $-models the dressing problem addresses the question:
What critical (d+1)-dimensional string is the gravitational dressed
version of what non-critical d-dimensional string?\\

The aim of this letter is the extension of the $\beta \leftrightarrow
\bar{\beta}$
relation to the next order of perturbation theory. Just the nextleading
order is still of considerable interest since at least in mass-independent
renormalization schemes for the simplest case of only one coupling the first
two orders are scheme independent. In section 2 we will show that for the
generic case eq.(\ref{2}) cannot be valid if higher orders are included.
This serves as a further motivation to look in section 3 for a formula
connecting
the nextleading orders of $\beta $ and $\bar{\beta}$.
\section{On the coupling constant dependence of the ratio $\beta$ to
$\bar{\beta}$}
For both $\beta$ and $\bar{\beta}$ we assume the perturbative structure
\footnote{The extension to perturbations including relevant operators
corresponding to linear terms in the $\beta$-functions is straightforward
\cite{Schmidh}, but will not be discussed further.}
\bea
\beta _{i}~=~\beta _{i}^{(2)ab}g_{a}g_{b}~+~\beta_{i}^{(3)abc}g_{a}g_{b}g_{c}
{}~+~...
\nonumber \\
\bar{\beta _{i}}~=~\bar{\beta }_{i}^{(2)ab}g_{a}g_{b}~+~\bar{\beta
}_{i}^{(3)abc}
g_{a}g_{b}g_{c}
{}~+~...~~.
\label{5}
\eea
Up to now it has been proven that
\beq
\bar{\beta}_{i}^{(n)}~=~\frac{2}{\alpha Q}~\beta _{i}^{(n)}~~\mbox{for}~~n=2~.
\label{6}
\eeq
We want to argue that eq.(\ref{6}) cannot be extended to all $n$.
For this purpose we assume the existence of a theory for which the RG fixed
point at $g_{i}~=~0$ is connected to a second RG fixed point at $g_{i}~=~
g_{i}^{\star}$ via a RG trajectory ($\beta _{i}\neq 0$ for $0<g_{i}<g_{i}^
{\star}$). Then
\beq
S_{c_{\star}}~=~S_{c}~+~\sum _{i} g_{i}^{\star}\int V_{i}~d^{2}z
\label{6a}
\eeq
describes a conformal theory with central charge $c_{\star}$. Due to
the c-theorem \cite{Zamo} $c_{\star}$ differs from $c$
\beq
c_{\star}~\neq c~.
\label{7}
\eeq
If we further assume $\left.\frac{\partial \beta_{i}}{\partial g_{j}}\right|
_{g=g^ {\star}}=0$ the expansion around $g=g^{\star}$ starts like that around
$g=0$ with quadratic terms. We can treat the problem symmetrically as a
perturbation around both points in coupling constant space.
If (\ref{2}) would be valid for the complete
$\beta $-func\-tions this would produce two different $\bar{\beta}$-functions:
$\frac{2}{\alpha Q} \beta $ and $\frac{2}{\alpha _{\star} Q_{\star}}\beta $
($\alpha _{\star},~Q_{\star}$ related to $c_{\star}$ in analogy to (\ref{3}),
(\ref{4})).\\

A universal factor relating $\beta $ to $\bar{\beta}$ can be interpreted simply
as a change in the renormalization scale $ \mu \rightarrow \bar {\mu}~=
{}~\mu^{\kappa}$ \cite{TKS}
\beq
\bar{\beta}(g)~=~\bar{\mu}\frac{\partial}{\partial \bar{\mu}}g~=~
\frac{\bar{\mu}}{\mu} \frac{\partial \mu}{\partial \bar{\mu}} \beta (g)
{}~=~\frac{1}{\kappa} \beta (g)~.
\label{8}
\eeq
Along this line of argument the difference between the two
$\bar{\beta}$-functions would be due to the use of different scales. One scale
is defined by means of the fixed point theory at $g~=~0$, the other
one by the fixed point theory at $g~=~g^{\star}$. Although this interpretation
looks similar to a RG-scheme dependence, it is completely unsatisfactory.\\
At every point of the RG trajectory $g \neq 0,~g\neq g^{\star}$
the gravitational dressed scale has to be defined just by means of the theory
under discussion. This requirement is of course fulfilled by the most natural
choice of scale given by the use of a cutoff in the geodesic distance.\\

Insisting on a unique $\bar{\beta}$-function
\footnote{understood modulo standard scheme dependence}
the lowest order result (\ref{6}) applied to the perturbation around both the
$c$
as well as the $c_{\star}$ theory states for this $\bar{\beta}$
\beq
\left.\frac{\partial ^{2} \bar{\beta}_{i}}{\partial g_{k}\partial g_{l}}\right|
_{g=0}~=~\left.
\frac{2}{\alpha Q}\frac{\partial ^{2}\beta_{i}}{\partial g_{k}\partial g_{l}}
\right| _{g=0}
{}~~~~\mbox{and}~~~~
\left.\frac{\partial ^{2} \bar{\beta}_{i}}{\partial g_{k}\partial g_{l}}\right|
_{g=g^{\star}}~=~
\left.\frac{2}{\alpha_{\star} Q_{\star}}\frac{\partial ^{2}\beta_{i}}{\partial
g_{k}\partial g_{l}} \right| _{g=g^{\star}}~~.  \label{9}
\eeq $\alpha Q \neq \alpha_{\star}Q_{\star}$ excludes an extension of (\ref{6})
to all $n$, i.e. the ratio $\frac{\bar{\beta}_{i}}{\beta_{i}}$ depends on $g$.
\section{Gravitational dressing of the nextleading order contribution to
$\beta$-functions}
We work in conformal gauge and use the method developed in refs.
\cite{Schmidh,AG}. It contains two steps. At first the gravitational dressed
action is written as
\beq
\tilde{S}~=~S_{c}~+~S_{L}~+~\sum_{i}g_{i}\int
\tilde{V}_{i}~\sqrt{\hat{g}}~d^{2}z~~.
\label{10}
\eeq
Here
\beq
S_{L}[\phi \vert \hat{g}]~=~\frac{1}{8\pi}\int d^{2}z\sqrt{\hat{g}}
\Big (\hat{g} ^{mn}\partial _{m}\phi \partial _{n} \phi + Q\hat{R}\phi (z)
+m^{2} e^{\alpha \phi} \Big )
\label{11}
\eeq
denotes the Liouville action for the Weyl degree of freedom of the
two-dimensional
metrics $g_{ab}=e^{\alpha \phi}\hat{g}_{ab}$. $S_{c}+S_{L}$ just describes the
gravitational dressed theory in the conformal case $g_{i}=0$ \cite{David,DK}.
For the dressing of the perturbations one makes the ansatz
\beq
\tilde{V}_{i}(z)~=~e^{\delta _{i} \phi (z)}~V_{i}(z)~~.
\label{12}
\eeq
Then the RG $\beta$-functions $\tilde{\beta}_{i}$ for this theory, understood
in the standard sense with respect to scaling in the coordinate $z$, are
calculated as an expansion in $g_{i}$ and $\delta _{i}$. The trace of the
corresponding energy-momentum tensor is given by (including the
contribution from diffeomorphism ghosts, $(\tilde{V}_{i})_{r}$ denoting
the renormalized dressed perturbations)
\beq
\tilde{T}_{a}^{a}~=~\frac{c-25+3Q^{2}}{24\pi}\hat{R}~+~\sum _{i}\tilde{\beta}_
{i}(\tilde{V}_{i})_{r}~~.
\label{13}
\eeq
Strictly speaking this formula can be derived for $m^{2}=0$ only. As is well
known the Liouville interaction cannot be treated as a perturbation, i.e.
it cannot be (with a corresponding $\tilde{\beta}$-function)
included in the sum of the r.h.s. easily. In analogy to the
handling of this problem in the case $g_{i}=0$ we assume that eq.(\ref{13})
remains valid for $m^{2}\neq 0$ if $\alpha$ is fixed by the requirement
that $e^{\alpha \phi}$ is a (1,1) operator. This may imply a dependence on
$g_{i}$, see below.\\
The vanishing of $\tilde{T}_{a}^{a}$, i.e. independence of the scale of
$\hat{g}_{ab}$ requires (\ref{3}) also for $g\neq 0$ as well as $\tilde{
\beta}_{i}~=~0$.
The last condition can be satisfied by treating the $\delta _{i}$ as
functions of the $g_{i}$.\\

After this construction of the gravitational dressed action $\tilde{S}$,
in a second step, one defines the gravitational dressed RG function
$\bar{\beta}
_{i}$ in connection to the response of a change of the cutoff in the
geodesic distance.
The procedure makes use of the relation between coefficients in
$\beta$-functions and operator product expansion (OPE) coefficients
\cite{Polyakov,Zamo,Cardy}. A direct extension to the next order has to
include the influence of the dressing by Liouville exponentials on the
OPE coefficients.\\

We found it convenient to use the arguments in a little bit implicit form
by studying the perturbative expansion of two-point functions. A similar
strategy has been used in refs.\cite{Lassig,CL}. For clarity we consider
the one coupling case only and leave the straightforward generalization to
several couplings for further work.\\
Correlation functions of $\tilde{V}$ with respect to the dressed unperturbed
action $S_{c}+S_{L}$ factorize in a conformal matter part and a Liouville
part. The scaling dimension of $V$ is two by assumption, that of $e^{\delta
\phi}$
is equal to $-y$, with $y$ defined by
\beq
y~=~\delta (\delta ~-~Q)~~.
\label{14}
\eeq
This leads to
\beq
\langle \tilde{V}(z_{1}) \tilde{V}(z_{2})\rangle _{0}~=~\frac{A_{2}(\delta)}
{\vert z_{1}-z_{2} \vert ^{4-2y}}~~,
\label{15}
\eeq
\beq
\langle \tilde{V}(z_{1})\tilde{V}(z_{2})\tilde{V}(z_{3})\rangle _{0}~=~
\frac{f\cdot A_{3}(\delta ,\delta ,\delta )}{(\vert z_{1}-z_{2}\vert
\vert z_{1}-z_{3}\vert \vert z_{2}-z_{3}\vert )^{2-y}}~~.
\label{16}
\eeq
The constant $f$ parametrizes the 3-point function of the undressed $V$, it
appears also in the OPE $V(z_{1})V(z_{2})=\frac{f}{\vert z_{1}-z_{2}\vert ^{2}}
V(z_{2})+...$\\
$A_{2}$ and $A_{3}$ are the $z$ independent factors in the 2 and 3-point
functions of Liouville exponentials \cite{DO2}. Using these building blocks
the full 2-point function
\beq
\langle \tilde{V}(z_{1})\tilde{V}(z_{2})\rangle~=~
\langle \tilde{V}(z_{1})\tilde{V}(z_{2})\rangle _{0}~-~
g\int d^{2} z \langle \tilde{V}(z_{1})\tilde{V}(z_{2})\tilde{V}(z)\rangle
_{0}~+~O(g^{2})
\label{17}
\eeq
becomes
\beq
\langle \tilde{V}(z_{1})\tilde{V}(z_{2})\rangle ~=~
\frac{A_{2}(\delta)}{\vert z_{1}-z_{2} \vert ^{4-2y}}~-~
\frac{\pi g f A_{3}(\delta ,\delta ,\delta )}{\vert z_{1}-z_{2}\vert ^{4-3y}}
\frac{\Gamma (1-y)(\Gamma (\frac{y}{2}))^{2}}{\Gamma (y)
(\Gamma (1-\frac{y}{2}))^{2}}~+~O(g^{2})~~.
\label{18}
\eeq
Obviously $y>0$ acts as a regulator. We now determine the renormalization
$Z$-factors as functions of $y$ and $g$. If $\mu$ is the RG-scale, then
the renormalized operator and coupling are defined by
\beq
\tilde{V}~=~Z_{\tilde{V}}\mu ^{-y}\tilde{V}_{r}
\label{19}
\eeq
\beq
g~=~Z_{g}\mu ^{y}g_{r}~~,
\label{20}
\eeq
respectively. From $\int d^{2} z\sqrt{\hat{g}}~\tilde{V}_{r}=\frac{\partial
\tilde{S}}
{\partial g_{r}}$ one finds the relation
\beq
Z_{\tilde{V}}^{-1}~=~Z_{g}~+~g_{r}\frac{\partial Z_{g}}{\partial g_{r}}~~.
\label{21}
\eeq
Out of $Z_{g}$ the wanted $\beta$-function $\tilde{\beta}$ is determined by
\cite{Amit,Lassig}
\beq
\tilde{\beta}~=~\frac{-y~g_{r}}{1~+~g_{r}\frac{\partial \log Z_{g}}
{\partial g_{r}}}~~.
\label{22}
\eeq
Using (\ref{21}) and
\beq
\frac{\Gamma (1-y)(\Gamma (\frac{y}{2}))^{2}}{\Gamma (y)
(\Gamma (1-\frac{y}{2}))^{2}}~=~\frac{4}{y}~(1~+~O(y^{2}))
\label{23}
\eeq
the renormalized 2-point function can be written as
\bea
\langle \tilde{V}(z_{1})\tilde{V}(z_{2})\rangle _{r}&=&\mu ^{2y}\frac{A_{2}}
{\vert z_{1}-z_{2}\vert ^{4-2y}}\left( Z_{g}+g_{r} \frac{\partial Z_{g}}
{\partial g_{r}}\right)^{2} \nonumber \\
\Big (1&-&(\mu \vert z_{1}-z_{2}\vert )^{y}g_{r}Z_{g}
\frac{4\pi f A_{3}}{yA_{2}} (1+O(y^{2}))~+~O(g_{r}^{2})\Big )~~.
\label{24}
\eea
Now $Z_{g}$ has to be choosen in a way to ensure finite r.h.s. for
$y\rightarrow
0$.\\

The most delicate point in this construction is the behaviour of $A_{2}$ and
$A_{3}$
for $\delta \rightarrow 0$.
Using the explicit formulas of ref. \cite{DO2}
\footnote{There is a change in notation. We call here the Liouville mass $m$
and the RG scale $\mu$.}
one gets a regular behaviour of $A_{2}(\delta )$ at $\delta = 0$ but
finds a divergence for $A_{3}(\delta ,\delta ,\delta )$ at $\delta =0$.
This pattern is connected with M\"obius invariance properties. To get a finite
2-point function for Liouville exponentials a division by the volume
$V^{(2)}_{CKV}$ of the subgroup of the M\"obius group leaving $z_{1}$ and
$z_{2}$ fixed is understood. This of course has to be done for all terms on the
r.h.s. of (\ref{18}), i.e. we must replace $A_{3}$ by $\frac{A_{3}}{V^{(2)}_
{CKV}}$. To give a meaning to this formal quantity, $V^{(2)}_{CKV}$ should be
regularized. However, since it is of no help to introduce a second independent
regularization, the only way to treat the problem rigorously would consist
in regularizing both $V^{(2)}_{CKV}$ and the usual UV-divergencies
\cite{Tseytlin,Polchinski} by means of a regularized propagator for the
Liouville field.
Such a calculation has not been done. Nevertheless we can expect a finite
answer for $\frac{A_{3}}{V^{(2)}_{CKV}}$. Altogether, taking into account a
proper normalization of $V^{(2)}_{CKV}$ we replace $\frac{A_{3}}{A_{2}}$ in
(\ref{24})
by $1+O(\delta )$.\\

Writing $Z_{g}$ in the form
\beq
Z_{g}~=~1~+~bg_{r}~+~O(g_{r}^{2})~~
\label{25}
\eeq
finite r.h.s. of (\ref{24}) up to $O(g_{r}^{2})$ is realized if
$$\Big (1+4bg_{r}+O(g_{r}^{2})\Big )\Big (1-\frac{4\pi f}{y} (1+O(y))
g_{r}+O(g_{r}^{2})\Big )$$
is finite up to $O(g_{r}^{2})$. From this we conclude
\beq
b~=~\frac{\pi f}{y}
\label{26}
\eeq
in a minimal subtraction scheme. In connection with (\ref{25}) and (\ref{22})
this yields
\beq
\tilde{\beta}(y,g_{r})~=~-y~g_{r}~+~\tilde{\beta}^{(2)}(y)~ g_{r}^{2}~+~
\tilde{\beta}^{(3)}(y)~g_{r}^{3}~+~O(g_{r}^{4})
\label{27}
\eeq
where
\beq
\tilde{\beta}^{(2)}(y)~=~\pi f~=~\beta ^{(2)}~~.
\label{28}
\eeq
In writing the second equation in (\ref{28}) we have taken into account
$\tilde{\beta}(0,g_{r})~=~\beta (g_{r})$ which is due to the decoupling
of the Liouville field for $y=0$.\\

To handle the nextleading gravitational dressing problem it is sufficient
to know from (\ref{28}) the $y$-independence of $\tilde{\beta}^{(2)}$ and
to assume regularity in $y$ for $\tilde{\beta}^{(3)}$ i.e.
\footnote{Similar to dimensional regularization one could expect
independence of $y$ for all $\tilde{\beta}^{(n)},~n\geq 2$. However, for us it
is
sufficient to assume the weaker property (\ref{29}).}
\beq
\tilde{\beta}^{(3)}(y)~=~\beta ^{(3)}~+~O(y)~~.
\label{29}
\eeq
Then the requirement $\tilde{\beta}=0$ determines
\beq
y~=~\beta ^{(2)}g_{r}~+~\beta ^{(3)} g_{r}^{2}~+~O(g_{r}^{3})~~,
\label{30}
\eeq
which via (\ref{14}) implies for $\delta $
\beq
\delta ~=~-\frac{\beta ^{(2)}}{Q}g_{r}~+~\frac{1}{Q}\left(\frac{(\beta ^{(2)})
^{2}}{Q^{2}}-\beta ^{(3)} \right)g_{r}^{2}~+~O(g_{r}^{3})~~.
\label{31}
\eeq
Let us turn to the cutoff regularized version of $\tilde{S}$ and denote the
cutoff in geodesic distance by $a$. It appears in the combination
$a^{2}\cdot e^{-\alpha \phi}$ only. This expression is proportional
to the squared cutoff $l^{2}$ in coordinate space, i.e.
\footnote{In this relation $\phi$ has to be understood as the Liouville
integration variable in the functional integral.}
\beq a^{2}~=~\hat{g}_{n}^{n}~l^{2}~e^{\alpha \phi}~~.
\label{32}
\eeq
A variation of $a$ can be compensated by a shift of the $constant$ mode
of the Liouville field related by
\beq
2~\frac{da}{a}~=~\alpha ~d\phi~~.
\label{33}
\eeq
This shift influences the linear term in $S_{L}$, the Liouville interaction
term
$m^{2}e^{\alpha \phi}$ and the dressed interaction term $g\tilde{V}$. The
linear term, due to the Gau{\ss}-Bonnet theorem, causes an overall factor
in front of the functional integral which drops out if normalization by
the partition function is taken into account. The Liouville interaction term
remains unchanged
if the shift of the Liouville field is compensated by a change of the
Liouville mass $m$
\beq
2~\frac{dm}{m}~=~-\alpha ~d\phi~~.
\label{34}
\eeq
Finally, requiring invariance also for the dressed interaction term yields
information on the dependence of the bare coupling $g$ on $m$ and $a$ at
fixed $g_{r}$
$$0~=~\left.d\big ( ge^{\delta \phi}V \big )\right|_{g_{r}}~=~
\Big (\frac{\partial g}{\partial a}da+\frac{\partial g}{\partial m}dm+
g\delta d\phi \Big )e^{\delta \phi}V~~. $$
With (\ref{33}),(\ref{34}) this gives
\beq
\frac{m\partial g}{\partial m}~-~\frac{a\partial g}{\partial a}~=~
\frac{2g\delta}{\alpha}~~.
\label{35}
\eeq
The solution of this equation is
\beq
g~=~\left(\frac{m}{\bar{\mu}}\right)^{\frac{\delta}{\alpha}}(a\bar{\mu}
)^{-\frac{\delta}
{\alpha}}f(am)g_{r}~~,
\label{36}
\eeq
$\bar{\mu}$ is a RG-scale, necessary for dimensional reasons.\\
The limit $a\rightarrow 0$ exists by construction since $y>0$ acts as
an effective regularization parameter.
\footnote{$y$ is fixed by (\ref{30}) already. This seems to restrict the
method to the subset of theories yielding positive $y$. However, without any
doubt the final result is valid for both signs of $y$.}
This fixes
$f(am)~=~(am)^{\frac{\delta}{\alpha}}$, i.e.  \beq g~=~\Big
(\frac{m}{\bar{\mu}}\Big )^{\frac{2\delta (g_{r})}{\alpha}}~g_{r}~~.
\label{37}
\eeq The gravitational dressed $\beta $-function is now defined by \beq
\bar{\beta}~ =~\left.m~\frac{\partial g_{r}}{\partial m}\right|
_{\bar{\mu},g~fix}~~.  \label{38} \eeq With (\ref{31}) this yields
\bea
\bar{\beta}(g_{r},\frac{m}{\bar{\mu}})&=&\frac{2\beta ^{(2)}}{\alpha Q}~
g_{r}^{2}~+~\frac{2}{\alpha Q}\left(\beta ^{(3)}-\frac{(\beta ^{(2)})^{2}}
{Q^{2}}\right)~g_{r}^{3}\nonumber \\ &+&\frac{4(\beta ^{(2)})^{2}}{\alpha
^{2}Q^{2}}~g_{r}^{3}~ \log \frac{m}{\bar{\mu}}~+~O(g_{r}^{4})~~.  \label{39}
\eea We still have to consider a possible $g_{r}$ dependence of $\alpha $.
As discussed in connection with the trace formula (\ref{13}) $\alpha $ should
be
fixed by requiring scaling dimension 2 for $e^{\alpha\phi}$ also in the case
$g\neq 0$. If one studies the two-point function  of this operator in analogy
to
that of $\tilde{V}$, due to $\langle \tilde{V}\rangle _{0}=0$, the perturbative
corrections start with $O(g^{2})$. Therefore, a $g_{r}$-dependence of $\alpha $
can influence $\bar{\beta}$ in order $O(g_{r}^{4})$ only.\\

The dependence of $\bar{\beta}$ on $\frac{m}{\bar{\mu}}$ fits into the
standard situation for $\beta $-functions in massive theories. By a
mass dependent redefinition of $g_{r}$
one can switch to a RG scheme with a mass independent $\bar{\beta}$-function.
In
the class of mass independent schemes $\bar{\beta}^{(3)}$ is unambiguously
given by
\beq
\bar{\beta}^{(3)}~=~\frac{2}{\alpha Q}\left( \beta ^{(3)}~-~\frac{(\beta ^{(2)}
)^{2}}{Q^{2}}\right)~~.
\label{40}
\eeq
A last comment concerns the overall sign of $\bar{\beta}$. Replacing
$m\frac{\partial }{\partial m}$ in (\ref{38}) by the more conventionally
looking $\bar{\mu}\frac{\partial}{\partial \bar{\mu}}$ would result in an
overall minus sign. However, since the overall sign is fixed by
the quasiclassical limit $(Q\rightarrow \infty ,~~\alpha Q\rightarrow 2)$
this sign cannot be accepted. One should mention that also the
dressed dimensions in the KPZ relation are equal to the power of $m$
and the negative power of $\bar{\mu}$ ( see \cite{David,DK} and
second ref. of \cite{DO1}).
\section{Concluding remarks}
Looking at our result (\ref{40}) from a more thorough point
of view the formal treatment of the M\"obius divergency appears as a weak spot
in its derivation. Further work should improve the status of the argument.
In parallel the result can be tested by calculations in concrete models.
One candidate for the simple one-coupling case is the Gross-Neveu model
coupled to gravity. The undressed $\beta $-function is known up to $O(g^{4})$
\cite{GN}. Other tests or applications are connected with the study of
continuum
limits of 2D lattice models on random lattices versus standard lattices.
In addition one should search for general relations between dressed and
undressed $\beta $-functions for generalized $\sigma $-models.\\[20mm]
\noindent
{\bf Acknowledgement}\\
I thank J. Ambjorn, D. L\"ust, H.-J. Otto and C. Schmidhuber for discussions
as well as J.L. Cardy for pointing out ref. \cite{Lassig}.


\end{document}